# Universal domain wall dynamics under electric field in Ta/CoFeB/MgO devices with perpendicular anisotropy

Weiwei Lin[1]*, Nicolas Vernier[1], Guillaume Agnus[1], Karin Garcia[1], Berthold Ocker[2], Weisheng Zhao[1], Eric E. Fullerton[3] and Dafiné Ravelosona[1*]

Electric field effects in ferromagnetic/oxide dielectric structures provide a new route to control domain wall (DW) dynamics with low power dissipation. However, electric field effects on DW velocities have only been observed so far in the creep regime where DW velocities are low due to strong interactions with pinning sites. Here, we show gate voltage modulation of DW velocities ranging from the creep to the flow regime in Ta/Co$_{40}$Fe$_{40}$B$_{20}$/MgO/TiO$_2$ structures with perpendicular magnetic anisotropy. We demonstrate a universal description of the role of applied electric fields in the various pinning dependent regimes by taking into account an effective magnetic field being linear with the electric field. In addition, the electric field effect is found to change sign in the Walker regime. Our work opens new opportunities for the study and optimization of electric field effect at ferromagnetic metal/insulator interfaces.

[1]Institut d'Electronique Fondamentale, Université Paris-Sud - CNRS, UMR8622, Orsay 91405, France, [2]Singulus Technologies AG, Kahl am Main 63796, Germany, [3]Center for Memory and Recording Research, University of California San Diego, La Jolla,

California 92093, USA. *email: linwei613@gmail.com (W.L.), dafine.ravelosona@u-psud.fr (D.R.).

Electric field control of magnetic domain wall (DW) motion in transition-metal ferromagnets[1–7] has attracted great interest due to the possibility to achieve DW control with low power dissipation for memory and logic circuits. One important example is racetrack memory[8], where a gate voltage can be exploited to assist manipulation of DWs under spin-polarized currents[1]. This would allow the reduction of the generally large critical currents needed to depin and shift DWs in narrow wires. Another example concerns DW logic circuits[9], for which local modulation of DW motion can lead to programmable DW energy landscapes. In both cases, the key is whether the electric field enables both efficient depinning for stored DWs and control of DW propagation.

Materials with perpendicular magnetic anisotropy (PMA) are particularly attractive for DW devices as they exhibit narrow DWs[10] enabling high-density storage. Recently, the electric field effect on DW dynamics has been demonstrated in Pt/Co/oxide films with PMA involving oxides with relatively high dielectric constant such as $AlO_x$[3,6], $HfO_2$[4] or $GdO_x$[5]. However, significant modulation of DW velocity under gate voltage has been shown only in the creep regime where DWs propagate at relatively low speeds (< 0.1 m s$^{-1}$) through thermal activation over local energy barriers[3–6] originating from the random disorder.[10,11] The efficient gate-voltage control in the creep regime has been explained in terms of modulation of the activation energy barriers[3–6] through voltage-induced changes of the interfacial magnetic anisotropy.[12–16] With increasing DW velocity in the creep regime, the electric field effect was found to be strongly reduced due to the decrease of the activation energies with applied magnetic fields.[5] So far, there has been no reports on the experimental observation of electric-field effects beyond the creep regime, particularly in the depinning and flow regimes.[17–20]

Here, we report the first observation of voltage modulation of DW motion over different dynamical regimes where velocities range from 10$^{-8}$ m s$^{-1}$ to 20 m s$^{-1}$. In addition, we demonstrate that this behavior can be understood over the full range of DW dynamics by electric field induced modulation of the magnetic anisotropy. Our approach makes use of Ta/CoFeB/MgO films with PMA, for which the density of pinning sites and the damping are very low[21-23] with respect to other PMA materials such as Co/Pt[19,24] or Co/Ni[18,25]. In addition, PMA at the CoFeB/MgO interface can be manipulated efficiently through gate-voltage control.[15,16] These materials are considered as the most promising not only for spin transfer torque magnetic random access memory (STT-MRAM)[26,27] but also for DW based memories[28], since a combination of spin Hall effect and Dzyaloshinskii-Moriya interaction leads to efficient DW propagation under current.[29–31]

In this study, DW velocity under electric field was measured in both as-grown and annealed Ta(5nm)/$Co_{40}Fe_{40}B_{20}$(1nm)/MgO(2nm)/$TiO_2$(20nm)/ITO samples using magneto-optic Kerr microscopy under combined electric and magnetic fields. By applying a top gate voltage $V_G$ between the bottom Ta/CoFeB and top ITO electrodes (shown schematically in Fig. 1a), electrons can either accumulate or deplete at the CoFeB/MgO interface. For the experiments shown, positive (negative) voltage corresponds to electron accumulation (depletion) at the CoFeB/MgO interface. Note that in order to avoid any temperature rise, the experiments were performed under low leakage current below 20 nA, which corresponds to gate voltages ranging from -1.5 V to 1.5 V. The maximum voltage corresponds to electric field of 0.37 V $nm^{-1}$ below the voltage breakdown of MgO (> 1 V $nm^{-1}$). Figure 1b shows typical differential Kerr images of magnetic DW motion in an annealed CoFeB structure for different gate voltages of $V_G$= -1.5 V, 0 V, +1.5 V. The dark regions indicate the motion of the DW under a magnetic field pulse of $\mu_0 H$ = 8 mT during $\Delta t$ = 1 s under voltage. The voltage $V_G$ was applied in the region of the 50-µm-wide ITO strip as indicated by the dashed rectangle. The DW displacement below the ITO strip at $V_G$ = −1.5 V is smaller than that at $V_G$ = 0 V, while the DW displacement at $V_G$ = 1.5 V is larger indicating an opposite effect of electric field for different polarities. Figures 2a-2c shows the dependence of DW velocity on the gate voltage in the annealed sample for different value of $\mu_0H$. In all field ranges, the DW velocity increases (decreases) with positive (negative) voltage $V_G$. As indicated in Fig. 2d, the effect of gate voltage on DW velocity is found to be relatively large in the low-field regime but it strongly decreases with increasing magnetic field.

To determine the role of voltage on the various dynamic regimes, we have measured the dependence of DW velocity $v$ on the applied magnetic field $H$ for various voltages ranging from -1.5 V to 1.5V. Figure 3 shows the DW velocity vs. $\mu_0H$ for typical gate voltages $V_G$= −1.5 V, 0 V and 1.5 V for the annealed sample. Three DW dynamical regimes are observed including the creep, intermediate depinning and depinning regimes[11,17,19,20,22]. In all regimes the DW velocity increases (decreases) under positive (negative) voltage. In the creep regime where $\mu_0H$ < 8 mT (Fig. 3b), the DW velocity can be expressed as

$$v^{\text{creep}} = v_0^{creep} \exp\left[-\frac{U_\text{C}}{k_\text{B}T}\left(\frac{H}{H_{\text{dep}}}\right)^{-\frac{1}{4}}\right] \quad (1)$$

where $U_\text{C}$ is a characteristic energy related to the disorder-induced pinning potential, $k_\text{B}$ the Boltzmann constant, $T$ the temperature, and $H_{\text{dep}}$ the depinning field at which the Zeeman energy is equal to the DW pinning energy. The exponent 1/4 fits the data well and is theoretically predicted for interactions of one-dimensional DWs with two-dimensional weak random disorder in thin magnetic films with PMA.[11] This regime allows us to determine the values of $U_\text{c}(H_{\text{dep}})^{-1/4}/k_\text{B}T$ and $\ln v_0^{creep}$ as a function of the applied voltage as shown in Figs. 4a-b. $U_\text{c}(H_{\text{dep}})^{-1/4}/k_\text{B}T$ decreases (increases) with positive (negative) voltages values whereas $\ln v_0^{creep}$ remains fairly constant. In agreement with Ref. 19, close above $H_{\text{dep}}$ a critical depinning regime is observed for $\mu_0H$ > 12 mT, where the velocity fits as

$$v^{\text{dep}} \sim \left(H - H_{\text{dep}}\right)^{\frac{1}{4}} \quad (2)$$

This regime allows us to determine accurately the value of $H_{\text{dep}}$ as a function of the gate voltages as shown in Fig. 4c. $H_{dep}$ decreases (increases) with positive (negative) voltage, with values ranging from $\mu_0 H_{dep}$= 11 mT at $V_G$ = +1.5 V to $\mu_0 H_{dep}$= 13 mT at $V_G$ = -1.5 V as compared to $\mu_0 H_{dep}$= 11.7 mT at $V_G$ = 0 V. We find a linear variation of the depinning field with voltage that can be described by

$$H_{dep} = H_{dep0}(1 - LV_G) \qquad (3)$$

where $H_{dep0}$ = 12 mT and $L \sim 0.05$ V$^{-1}$. Finally, by using the value of $\mu_0 H_{dep}$ determined from Eq. (2) into Eq. (1), the characteristic energy $U_c$ is found to be independent of the voltage as indicated in Fig. 4d. Between these two regimes for 8 mT < $\mu_0 H$ < 12 mT, an intermediate depinning regime occurs, which corresponds to the tails of the creep regime[20] where the energy barriers vanishes linearly as $\Delta E \sim (H/H_{dep}-1)$ approaching $H_{dep}$.

Our results are consistent with electric field induced change of perpendicular anisotropy as described below. In our structure, the dielectric layer consists of a 2 nm thick MgO (dielectric constant $\varepsilon_{MgO} \sim 9.7$) and 20 nm thick TiO$_2$ (dielectric constant $\varepsilon_{TiO2} \sim 100$) films. A positive (negative) gate voltage of $V_G$ = +1 V (−1 V) corresponds to an electric field of ~ 0.25 V nm$^{-1}$ and an accumulation (depletion) of ~ 0.01 electron per Co(Fe) atom at the CoFeB/MgO interface. The results here show that positive voltage increases the DW velocity. This is consistent with recent studies showing that positive voltage give rise to a reduction of the effective anisotropy $K_{eff}$ in Ta/CoFeB/MgO structures[15,16,32-34]. Indeed, since the depinning field $H_{dep}$ depends on $K_{eff}$ as $H_{dep} \sim (K_{eff})^{1/2}$, a reduction of $K_{eff}$ under positive voltage leads to a reduction of $H_{dep}$ and thus an increase of DW velocity following Eqs.1 and 2. Note that the variation of $H_{dep}$ gives rise to a depinning efficiency of 2.3 mT per V nm$^{-1}$.

In a recent study[34], we have used an ionic liquid to apply electric fields on larger scales in Ta/CoFeB/MgO structures grown under the same conditions (see supplementary information), which allows us to measure the variation of the anisotropy field under electric field. The results are found to be consistent with a linear variation of the effective anisotropy under voltage confirming previous studies[16,32]. Thus considering $K_{eff} = K_{eff0}(1 - aV_G)$ where $K_{eff0}$ is the effective anisotropy at zero voltage, $a$ is the modulation ratio of $K_{eff}$ per V, and $V_G$ the gate voltage, the depinning field $H_{dep} \sim (K_{eff})^{1/2}$ can be written as

$$H_{dep} = H_{dep0}(1 - aV_G/2) \qquad (4)$$

where $H_{dep0}$ is the propagation field at zero voltage. This is consistent with our finding of a linear variation of $H_{dep}$ with voltage (see Eq. 3) considering $a/2 = L$. Using $L = 0.05$ V$^{-1}$ deduced from the fit to Fig. 4c (see also S7), we find that the modulation ratio of $K_{eff}$ is $a \sim 10\%$ per V. Since the modulation of $K_{eff}$ is related to the modulation of the interface anisotropy at the CoFeB/MgO interface, and $K_{eff0} \sim 10^5$ J/m$^3$ in our films, the typical modulation ratio of interface anisotropy $K_{eff}t_{CoFeB}$ where $t_{CoFeB} = 1$ nm is then 40 µJ/m$^2$ per V nm$^{-1}$ in agreement with previous studies on Ta/Co$_{40}$Fe$_{40}$B$_{20}$/MgO[15].

In the creep and intermediate depinning regime where DW velocities follow an exponential behavior, the energy barrier depends on $H/H_{dep}$ that can be written $H/H_{dep} = (1 + LV_G)H/H_{dep0}$. This indicates that the DW velocity under voltage follows the same expressions as the ones at zero voltage by replacing the magnetic field $H$ by an effective magnetic field $H_{eff} = (1 + LV_G)H$, i.e, the energy barrier is given by $H_{eff}/H_{dep0}$. This is shown in Fig. 5a where the DW velocity $v$ is plotted against $H_{eff}$ under typical $V_G$ of −1.5 V, 0 V and 1.5 V respectively. The behavior in the DW creep regime is also shown in Fig. 5b where $v$ (in logarithmic scale) is plotted against $(H_{eff})^{-1/4}$. All the curves can be superimposed by considering the effective magnetic field instead of the applied magnetic field. It is interesting to observe that it is also the case in the depinning regime above $H_{dep}$ where the DW velocity depends on $(H - H_{dep})^{1/4}$. Since $LV_G \ll 1$, $(H - H_{dep})^{1/4} \sim (H_{eff} - H_{dep0})^{1/4}$, which indicates that the universal description given by $H_{eff}$ is quantitatively applicable over the full range of the pinning dependent regimes.

Finally, Figure 6 shows DW velocity at high fields in an as-grown Ta(5nm)/Co$_{40}$Fe$_{40}$B$_{20}$(1nm)/MgO(2nm)/TiO$_2$(20nm)/ITO structure, for which the PMA is lower due to the amorphous character of the CoFeB layer.[19] In this case, the beginning of the flow regime occurs at lower fields than the annealed sample, which allows us to study the influence of electric field on the flow regime. As shown in Fig. 5a, the velocity $v$ at $V_G = 0$ as a function of the magnetic field increases up to $\mu_0 H = 27$ mT. Interestingly, for $\mu_0 H > 27$ mT, a new regime where the DW velocity decreases with increasing $H$ is observed. This additional regime beyond the depinning regime corresponds to the instability regime above the Walker breakdown that is partially hidden by the depinning regime as we have recently observed[17,18,21,31]. For $\mu_0 H < 27$ mT the DW velocity increases with positive $V_G$, as shown in Fig. 5b. Interestingly Fig. 5c shows that for $\mu_0 H = 30$ mT in the negative mobility regime, instead the DW velocity decreases with increasing $V_G$, indicating a reversal of the electric field effect. In this regime, the DW velocity exhibits a non-trivial dependence on the effective anisotropy. First, we note that the Walker field can be written $H_W = N_y M_S \alpha / 2$, where $N_y$ is the demagnetizing factor across the wall given as $t_{CoFeB}/(t_{CoFeB} + \Delta)$, $\alpha$ is the damping parameter and $\Delta$ is the domain wall width that can be written $\Delta = (A/K_{eff} + N_y \mu_0 M_S^2 /2)^{1/2}$. The typical value of $H_W$ is around 0.8 mT in our films[22] lower than $H_{dep}$. Since $H_W$ depends on $\Delta^{-1}$, $H_W$ is expected to vary as $(K_{eff})^{1/2}$ such as $H_{dep}$. Thus a positive voltage (reduction of $K_{eff}$) results in a decrease of $H_W$, which as first approximation translates the Walker regime to lower fields reducing the DW velocity above $H_W$ as we observe. This is further supported by micromagnetic simulations (see S6), which shows that in the Walker regime indeed a reduction of anisotropy gives rise to a reduction of domain wall velocity.

In conclusion, beyond the important finding of the universal description of electric field induced DW motion, the possibility to both obtain depining efficiency up to 2.5 mT per V nm$^{-1}$ and control DW velocity up to the flow regime in CoFeB-MgO structures opens new perspectives for low power spintronic applications such as solid state memories and logic devices. We believe that these electric field effects can be used to lower the energy barrier for stored DWs leading to a smaller spin polarized currents to depin and move them. Also, as the depinning fields of CoFeB-MgO can be as small as 2-3mT, On/Off operations are feasible, which is of interest for logic devices. Furthermore, electric field effects can be very useful in the flow regimes to fasten DW motion between two storing positions.

**Methods**

**Fabrication of Ta/CoFeB/MgO/TiO$_2$/ITO structure.** The Ta(5 nm)/Co$_{40}$Fe$_{40}$B$_{20}$(1 nm)/MgO(2 nm)/Ta(5 nm) films were deposited on Si/SiO$_2$ wafers by magnetron sputtering. Selected films were annealed at 300 °C for two hours. After etching the top 5 nm thick Ta layer, a 20 nm thick TiO$_2$ dielectric layer was sputtered on the Ta/CoFeB/MgO film (see S1). Finally, 100 nm thick sputtered ITO (In$_2$O$_3$:Sn) stripes with width ranging from 10 μm to 50 μm were patterned on top of the TiO$_2$ layer as the top electrode by optical lithography and lift-off process. ITO was used as the top electrode because it is not only a good electrical conductor (about 7 Ω$^{-1}$ m$^{-1}$ conductivity), but also is transparent for visible light, so that it is possible to use Kerr microscopy for visualizing magnetic domains. By applying a top gate voltage $V_G$ between the bottom Ta/CoFeB and the top ITO, electrons can accumulate or deplete at the CoFeB/MgO interface. Here, positive (negative) gate voltage corresponds to electron accumulation (depletion) in the CoFeB layer. The leakage current was measured to be less than a few nA for gate voltages $V_G$ < 1.5 V and less than 100 nA at the maximum gate voltage and thus, Joule heating influence on DW motion can be negligible.

**Magnetic domain wall velocity measurement by polar Kerr microscopy.** The polar Kerr microscopy was performed using a diode with the blue light (455 nm in wavelength) and a ×50 objective lens. The analyzer was set to about 87 degree (analyzer and polarizer almost crossed) and the exposure time was 2 s. To measure DW velocity under an applied perpendicular magnetic field $H$, we used the following procedures: first, the sample was saturated with a magnetic field of −20 mT fixed for more than 10 s. Second, several positive magnetic field pulses of +3 mT with duration of around 10 s were applied, until the nucleation of a single DW below the ITO strip was observed. A reference Kerr image was taken at zero applied magnetic field after the first set of magnetic field pulses. Third, another magnetic field pulse was applied resulting in a DW displacement. The amplitude of the second set of magnetic field pulses ranged from 0.6 mT to 35 mT and their durations Δ$t$ from 100 s (long pulse) to 9 µs (short pulse). The magnetic field pulse was generated from a coil with 1 cm diameter, which was fixed very close to the sample, and the rise time of magnetic field was about 4 µs. After the second set of magnetic field pulses were applied, a second Kerr image was taken as soon as the magnetic field was back to zero. The DW displacement during Δ*t* was obtained by the difference between the two Kerr images, which gave the DW velocity *v*. The velocity measured under the short magnetic field pulses was determined by the slope of the DW displacement vs. magnetic field pulse duration.

**References**

1. Yamanouchi, M., Chiba, D., Matsukura, F., Ohno, H. Current-assisted domain wall motion in ferromagnetic semiconductors. *Jap. J. Appl. Phys*. **45**, 3854–3859 (2006).
2. Logginov, A. S. *et al*. Room temperature magnetoelectric control of micromagnetic structure in iron garnet films, *Appl. Phys. Lett*. **93**, 182510 (2008).
3. Schellekens, A. J., van den Brink, A., Franken, J. H., Swagten, H. J. M. & Koopmans, B. Electric-field control of domain wall motion in perpendicularly magnetized materials. *Nature Commun*. **3**, 847 (2012).
4. Chiba, D. *et al*. Electric-field control of magnetic domain-wall velocity in ultrathin cobalt with perpendicular magnetization. *Nature Commun*. **3**, 888 (2012).
5. Bauer, U., Emori, S. & Beach, G. S. D. Voltage-gated modulation of domain wall creep dynamics in an ultrathin metallic ferromagnet. *Appl. Phys. Lett*. **101**, 172403 (2012).
6. Bernand-Mantel, A. *et al*. Electric-field control of domain wall nucleation and pinning in a metallic ferromagnet. *Appl. Phys. Lett*. **102**, 122406 (2013).
7. Bauer, U., Emori, S. & Beach, G. S. D. Voltage-controlled domain wall traps in ferromagnetic nanowires. *Nature Nanotechnol*. **8**, 411–416 (2013).
8. Parkin, S. S. P., Hayashi, M. & Thomas, L. Magnetic domain-wall racetrack memory. *Science* **320**, 190–194 (2008).

9. Allwood, D. A. *et al*. Submicrometer ferromagnetic NOT gate and shift register. *Science* **296**, 2003–2006 (2002).

10. Ferré, J. in *Spin Dynamics in Confined Magnetic Structures I*, edited by B. Hillebrands and K. Ounadjela (Springer-Verlag Berlin Heidelberg, 2002), Vol. 83, p. 127.

11. Lemerle, S. *et al*. Domain wall creep in an Ising ultrathin magnetic film. *Phys. Rev. Lett*. **80**, 849–852 (1998).

12. Weisheit, M. *et al*. Electric field-induced modification of magnetism in thin-film ferromagnets. *Science* **315**, 349–351 (2007).

13. Maruyama, T. *et al*. Large voltage-induced magnetic anisotropy change in a few atomic layers of iron. *Nature Nanotechnol*. **4**, 158–161 (2009).

14. Niranjan, M. K., Duan, C.-G., Jaswal, S. S. & Tsymbal, E. Y. Electric field effect on magnetization at the Fe/MgO(001) interface. *Appl. Phys. Lett*. **96**, 222504 (2010).

15. Endo, M. *et al*. Electric-field effects on thickness dependent magnetic anisotropy of sputtered MgO/Co$_{40}$Fe$_{40}$B$_{20}$/Ta structures. *Appl. Phys. Lett*. **96**, 212503 (2010).

16. Kita, K., Abraham, D. W., Gajek, M. J. & Worledge, D. C. Electric-field-control of magnetic anisotropy of Co$_{0.6}$Fe$_{0.2}$B$_{0.2}$/oxide stacks using reduced voltage. *J. Appl. Phys*. **112**, 033919 (2012).

17. Metaxas, P. J. *et al*. Creep and flow regimes of magnetic domain-wall motion in ultrathin Pt/Co/Pt films with perpendicular anisotropy. *Phys. Rev. Lett*. **99**, 217208 (2007).

18. Yamada, K. *et al*. Influence of instabilities on high-field magnetic domain wall velocity in (Co/Ni) nanostrips. *Appl. Phys. Express* **4**, 113001 (2011).

19. Gorchon, J. *et al*. Pinning-dependent field-driven domain wall dynamics and thermal scaling in an ultrathin Pt/Co/Pt magnetic film. Phys. Rev. Lett. **113**, 027205 (2014).

20. Jeudy, V et al. Universal pinning energy barrier for driven domain walls in thin ferromagnetic films, arXiv:1603.01674v1.

21. Devolder, T. *et al*. Damping of Co$_x$Fe$_{80-x}$B$_{20}$ ultrathin films with perpendicular magnetic anisotropy. *Appl. Phys. Lett*. **102**, 022407 (2013).

22. Burrowes, C. *et al*., Low depinning fields in Ta-CoFeB-MgO ultrathin films with perpendicular magnetic anisotropy. *Appl. Phys. Lett*. **103**, 182401 (2013).

23. Tetienne, J-P. et al., Nanosacle imaging and control of domain wall hoping with a nitrogen vacancy center microscope, *Science* **344**, 1366 (2014).

24. Mizukami, S. *et al*., Gilbert damping in perpendicularly magnetized Pt/Co/Pt films investigated by all-optical pump-probe technique. *Appl. Phys. Lett*. **96**, 152502 (2010).

25. Mizukami, S. *et al*., Gilbert damping in Ni/Co multilayer films exhibiting large perpendicular anisotropy. *Appl. Phys. Express* **4**, 013005 (2011).

26. Ikeda, S. *et al*. A perpendicular-anisotropy CoFeB-MgO magnetic tunnel junction. *Nature Mater*. **9**, 721–724 (2010).

27. Worledge, D. C. *et al*. Spin torque switching of perpendicular Ta/CoFeB/MgO-based magnetic tunnel junctions. *Appl. Phys. Lett*. **98**, 022501 (2011).

28. Fukami, S. *et al*. Current-induced domain wall motion in perpendicularly magnetized CoFeB nanowire. *Appl. Phys. Lett*. **98**, 082504 (2011).

29. Thiaville, A., Rohart, S., Jué, E., Cros, V. & Fert, A. Dynamics of Dzyaloshinskii domain walls in ultrathin magnetic films. *Europhys. Lett*. **100**, 57002 (2012).

30. Ryu, K.-S., Thomas, L., Yang, S.-H. & Parkin, S. Chiral spin torque at magnetic domain walls. *Nature Nanotechnol*. **8**, 527–533 (2013).

31. Emori, S., Bauer, U., Ahn, S.-M., Martinez, E. & Beach, G. S. D. Current-driven dynamics of chiral ferromagnetic domain walls. *Nature Mater*. **12**, 611–616 (2013).

32. Kanai, S., Martin Gajek, M., Worledge, M., Matsukura, F., Ohno, H., Electric field-induced ferromagnetic resonance in a CoFeB/MgO magnetic tunnel junction under dc bias voltages, Appl. Phys. Lett. **105**, 242409 (2014)

33. Wang, W. G., Li, M., Hageman, S. & Chien, C. L. Electric-field-assisted switching in magnetic tunnel junctions. *Nature Mater*. **11**, 64–68 (2012)

34. Liu, Y.T, Agnus, G., Ono, S., Ranno, L., Bernand-Mantel, A., Soucaille, R.,

Adam, J.P, Langer, J., Ocker, B., Ravelosona, D., and Herrera Diez, L., Ionic-liquid gating of perpendicularly magnetised CoFeB/MgO thin films. Under consideration in Appl. Phys. Lett (2016)

**Acknowledgements**

This work was partly funded by the European Community's Seventh Framework Programme FP7 under grant agreement 257707 (MAGWIRE); the French Agence Nationale de la Recherche (ANR) through the projects ELECMADE, FRIENDS, DIPMEN and COMAG; and by RTRA, C'Nano IdF and LABEX NanoSaclay. Work at UCSD supported by the National Science Foundation under award number DMR-1312750. We thank T. Devolder and S. Eimer for experimental assistance, J. P. Adam for micromagnetic simulation, J. V. Kim, L. Herrera Diez, L. Ranno, A. Marty, A. Bernand-Mantel, D. Givord, J. P. Jamet, A. Thiaville and V. Jeudy for fruitful discussions.

**Author contributions**

W.L. and D.R. designed the experiment. B.O. optimized the deposition of Ta/CoFeB/MgO/Ta film. W.L. fabricated the Ta/CoFeB/MgO/TiO$_2$/ITO structures with the help of G.A. and K.G. W.L. and N.V. performed polar Kerr microscopy measurement and analyzed the data. W.L. gave the phenomenological physics description and wrote the manuscript. N.V., E.E.F. and D.R. contributed to the manuscript revision. W.Z. discussed the work. D.R. managed the study.

**Competing financial interests**

The authors declare no competing financial interests.

**Figure captions**

**Figure 1 : Schematic of the experiment and magnetic domain wall displacement under gate voltages in annealed Ta/Co$_{40}$Fe$_{40}$B$_{20}$/MgO/TiO$_2$ structures. a** The gate voltage $V_G$ is applied between the ITO strip and the CoFeB film. The magnetic field **H** is applied perpendicular to the film plane. **b-d** DW displacement measured by polar Kerr microscopy at applied magnetic fields $\mu_0H$ of 8 mT with 1 s duration under voltage $V_G$ of −1 V, 0 V and 1 V, respectively. The DW displacement is obtained from the difference between the polar Kerr images after and before magnetic field pulses. The right (left) boundary in the dark part in the image shows the DW position before (after) applying the magnetic field pulse, and the arrow indicates the direction of DW motion. Top gate voltage $V_G$ is applied in the region indicated by the dashed rectangle.

**Figure 2 : Magnetic DW velocity as a function of the voltage gate for different magnetic fields. a-c**, Domain wall velocity as a function of $V_G$ for different magnetic fields $\mu_0H$ of 3 mT, 9 mT and 29 mT, respectively. (d) Relative variation of DW velocity v(-1.5V)-v(+1.5V)/v(0V) as a function of the applied field. Electric field effect is detect up to the high velocity regime.

**Figure 3 : Magnetic DW velocity as a function of the applied magnetic field for different gate voltages in annealed Ta/CoFeB/MgO structures. a,** *velocity* as a function of $H$ under voltages $V_G$ of −1.5 V (open squares), 0 V (solid squares) and 1.5 V (open circles). **b,** *velocity* (in logarithmic scale) as a function of $H^{-1/4}$ under voltages $V_G$ of −1.5 V (open squares), 0 V (solid squares) and 1.5 V (open circles). The symbols correspond to the experimental data, and the lines to the fitting (see text).

**Figure 4 : Analysis of the creep regime under voltages. a-d** Variation of $U_c(H_{dep})^{-1/4}/k_BT$, $\mu_0H_{dep}$, $lnv_0$ and $U_c/k_BT$ respectively as a function of the gate voltage.

**Figure 5 : Domain wall velocity as a function of the effective magnetic field Heff for different gate voltages in annealed Ta/CoFeB/MgO structures. a,** Velocity *v* as a function of $(1+LV_G)H$ under voltages $V_G$ of −1.5 V (open squares), 0 V (solid squares) and 1.5 V (open circles), where $L$ = 0.05 V$^{-1}$. **b,** Velocity *v* (in logarithmic scale) as a function of $[(1+LV_G)H]^{-1/4}$ under voltages $V_G$ of −1.5 V (open squares), 0 V (solid squares) and 1.5 V (open circles), where $L$ = 0.05 V$^{-1}$.

**Figure 6 : Gate voltage effect on magnetic DW velocity in an as-deposited Ta/CoFeB/MgO structure. a,** Velocity *v* as a function of *H* at zero voltage. **b,c,** *Domain wall* velocity *v* as a function of $V_G$ under magnetic field $\mu_0 H$ of 6 mT and 30 mT, respectively.

**Figure 1**

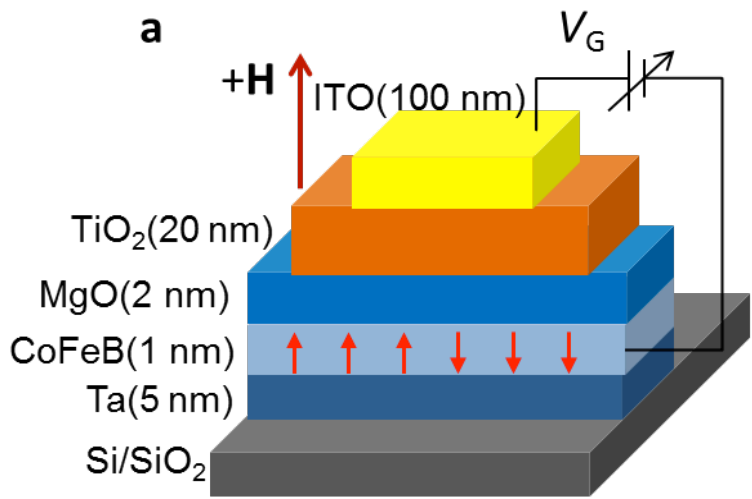

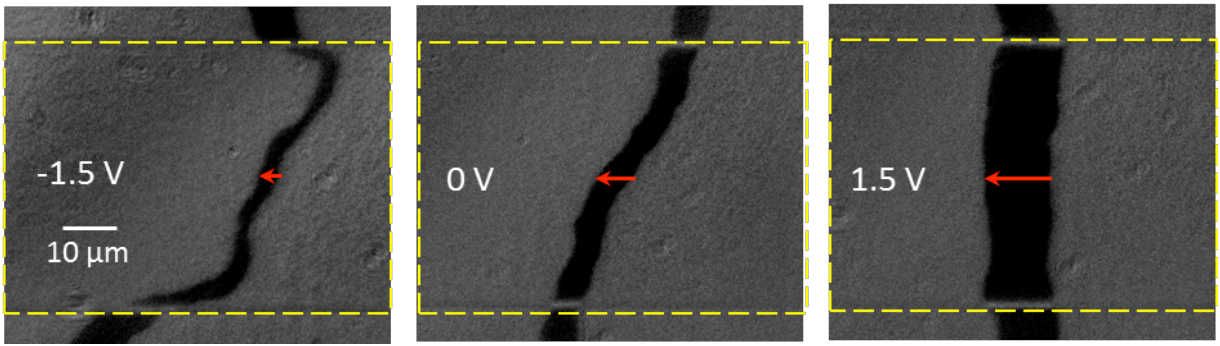

**b** $\mu_0 H = 8$ mT

-1.5 V  |  0 V  |  1.5 V

10 μm

# Figure 2

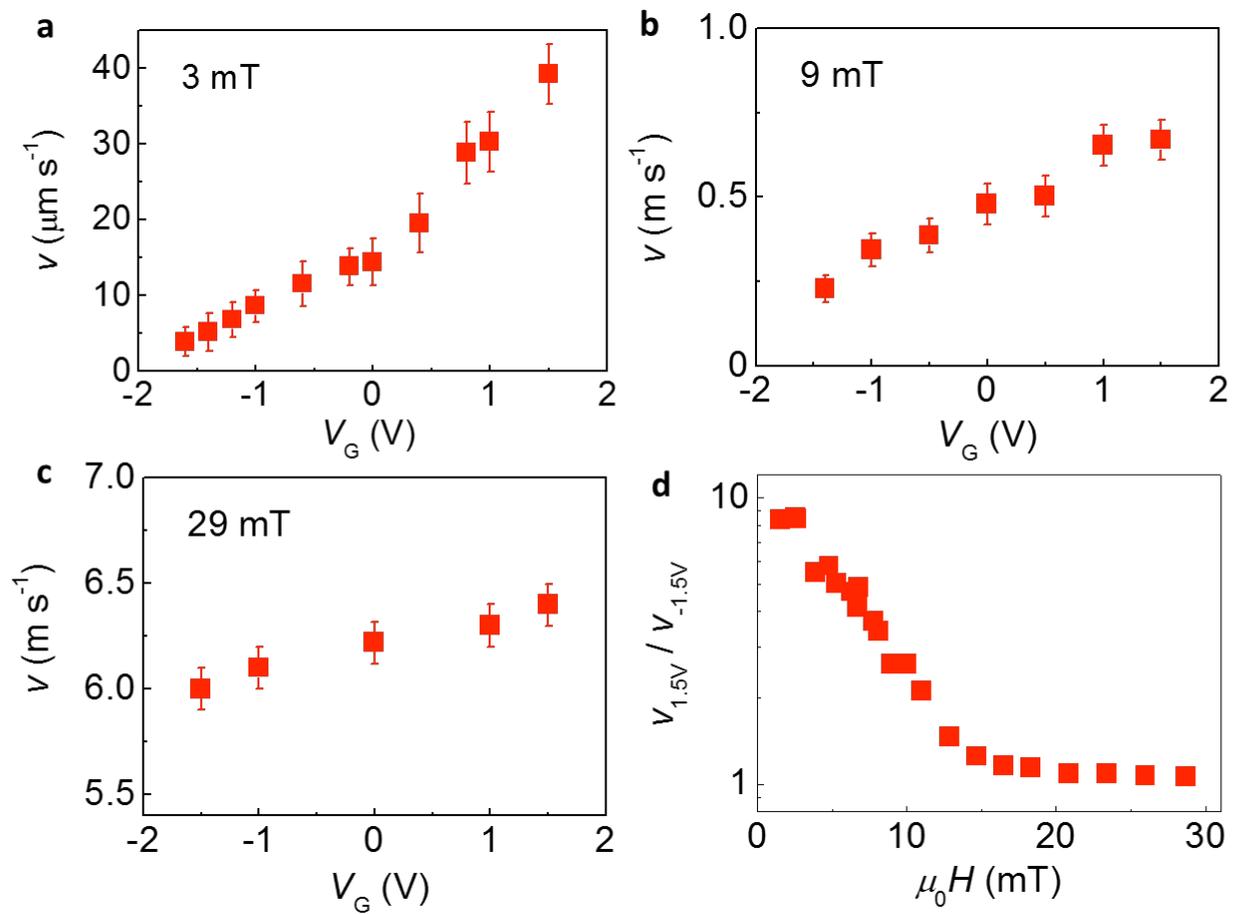

**Figure 3**

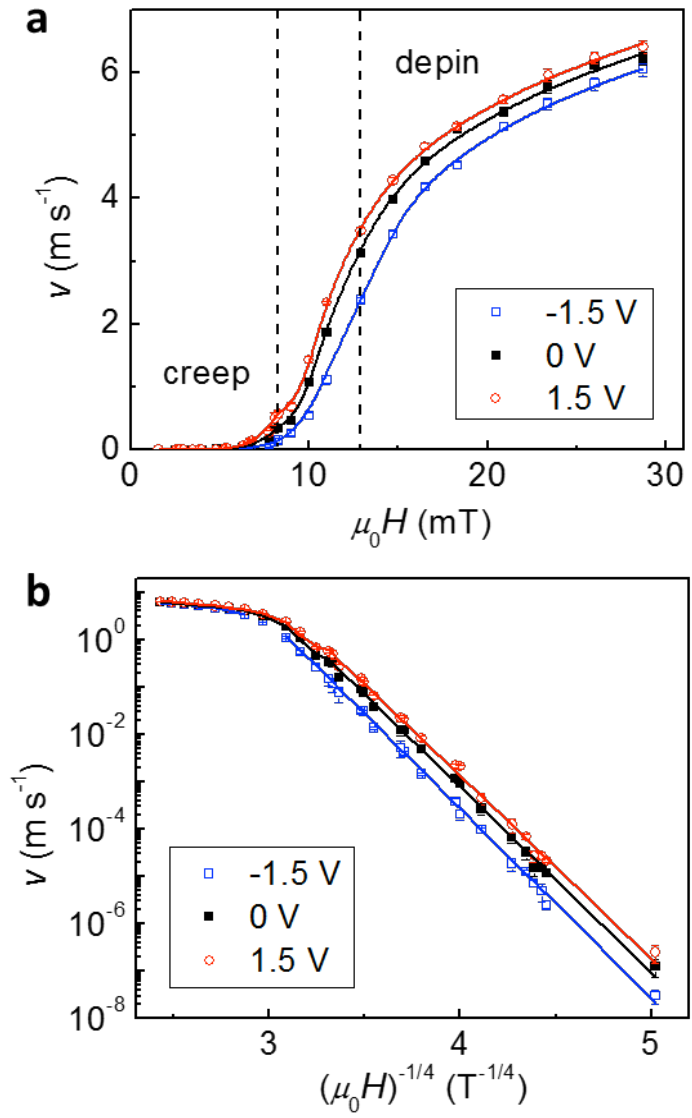

**Figure 4**

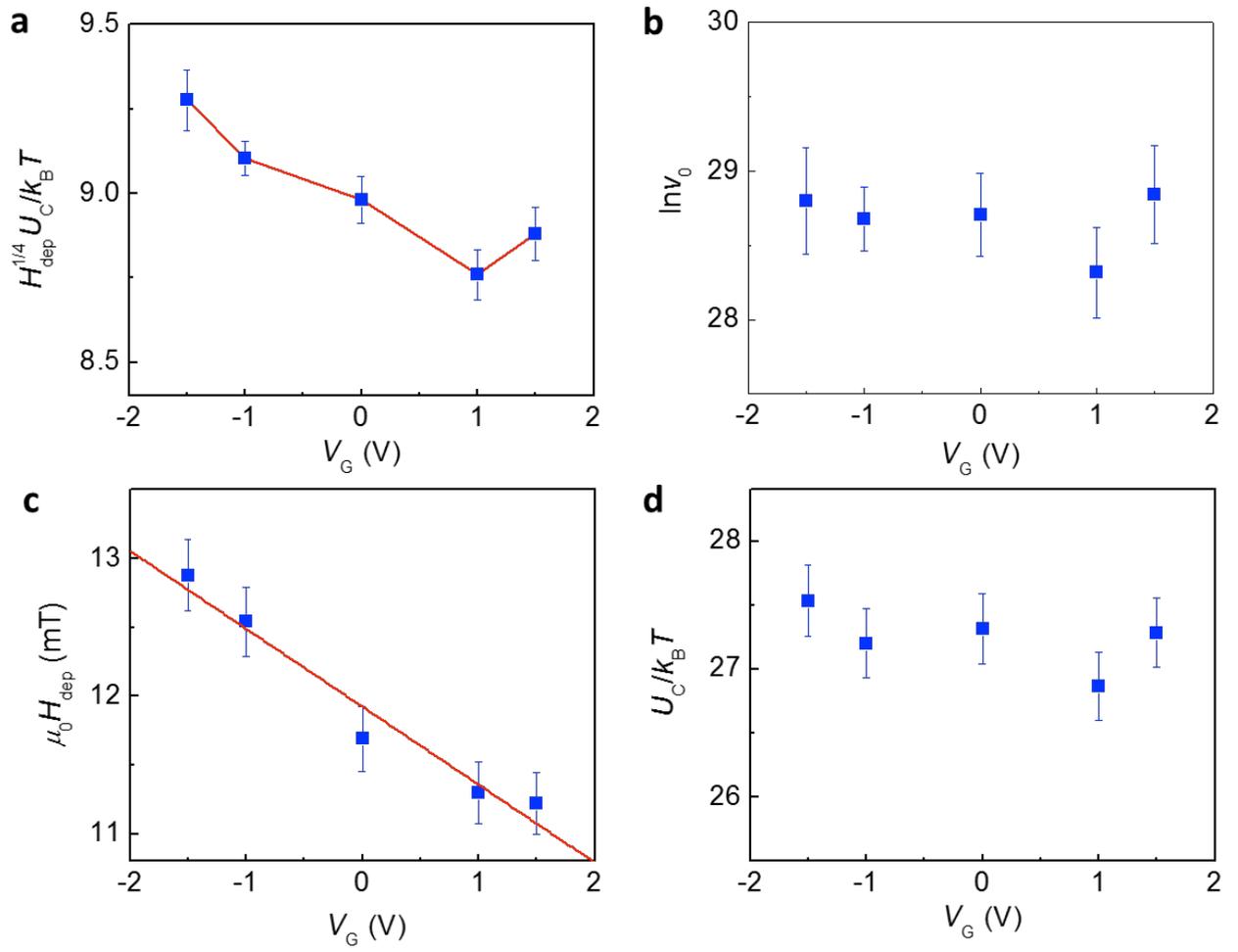

**Figure 5**

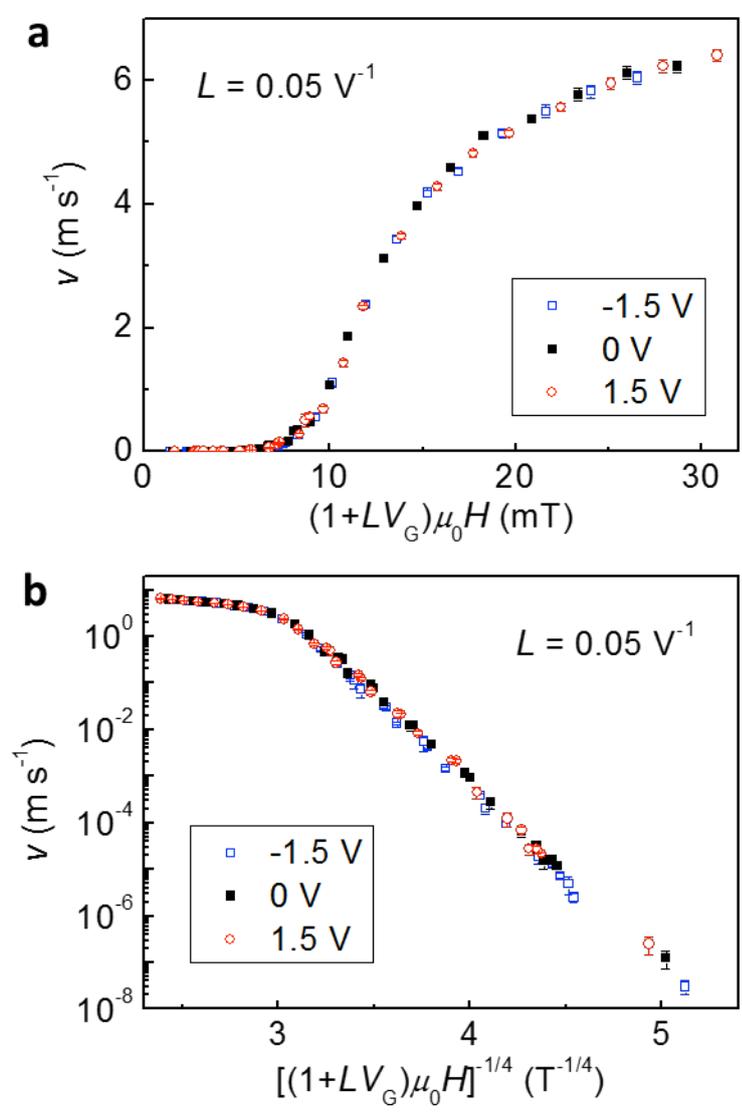

**Figure 6**

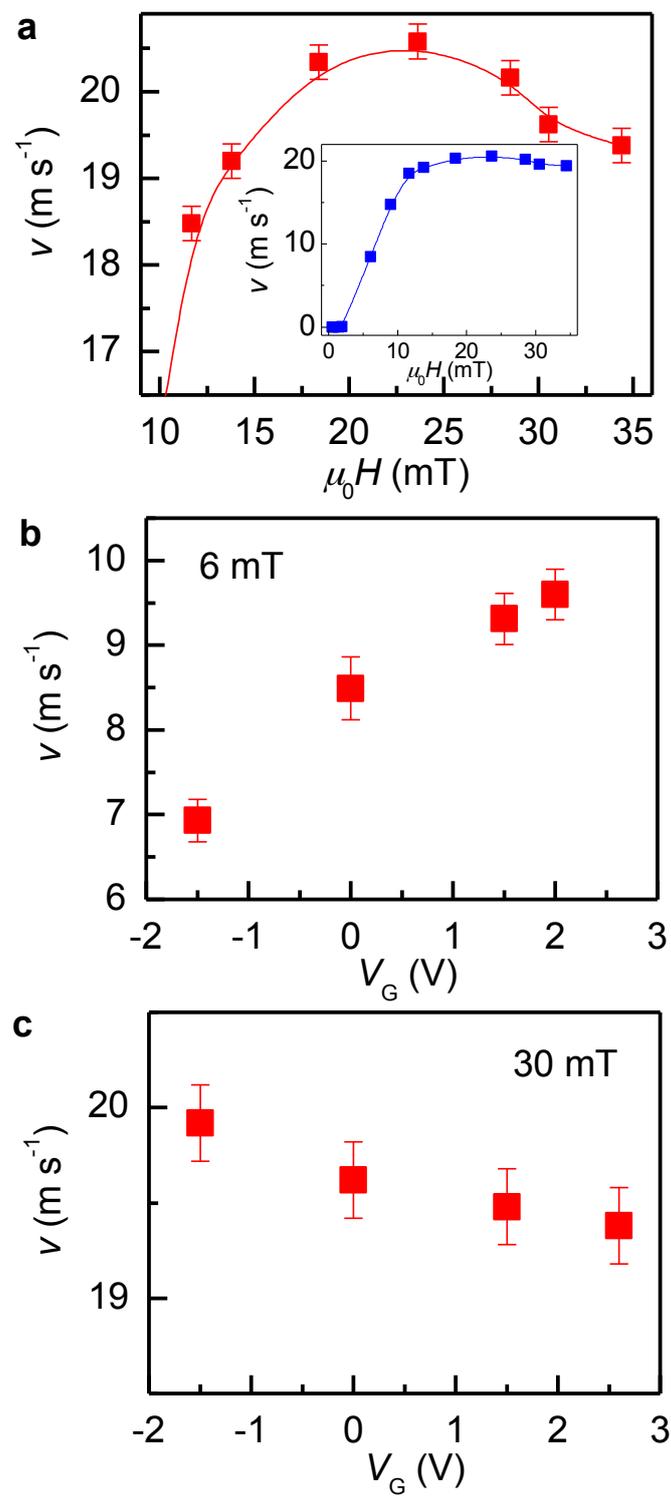